\documentclass{aa}  

\usepackage{graphicx}
\usepackage{txfonts}
\usepackage{lipsum}
\usepackage{subcaption}         
\usepackage{lscape}             
\usepackage{placeins}           
\makeatletter
\let\linenumbers\relax
\let\@@par\@@@par
\makeatother

\begin{document}

   \title{Upstream neutrino production and delayed jet emission in the blazar GB6 J1542+6129}

\author{
Emma Kun\inst{1,2,3,4}\corrauth{kun.emma@csfk.org}
\and Imre Bartos\inst{5}\email{imrebartos@ufl.edu}
\and Breshna Hadi\inst{6}
\and Anna G\"obly\"os\inst{7}
\and Julia Becker Tjus\inst{4,8,9}
\and Peter L.\ Biermann\inst{10,11}
\and Anna Franckowiak\inst{6}
\and Francis Halzen\inst{12}
\and Santiago del Palacio\inst{8}
\and Claudio Ricci\inst{13, 14}
}

\institute{
Department of Astronomy, Institute of Physics and Astronomy, ELTE, P\'azm\'any P\'eter s\'et\'any 1a, H-1117 Budapest, Hungary
\and Konkoly Observatory, HUN-REN Research Centre for Astronomy and Earth Sciences, Konkoly Thege Mikl\'os \'ut 15--17, H-1121 Budapest, Hungary
\and CSFK, MTA Centre of Excellence, Konkoly Thege Mikl\'os \'ut 15--17, H-1121 Budapest, Hungary
\and 
Theoretical Physics IV, Faculty for Physics \& Astronomy, Ruhr University Bochum, 44801 Bochum, Germany
\and Department of Physics, University of Florida, PO Box 118440, Gainesville, FL 32611-8440, USA
\and Astronomical Institute, Faculty for Physics \& Astronomy, Ruhr University Bochum, 44780 Bochum, Germany
\and Institute of Physics, University of Szeged, Szeged, Hungary
\and Ruhr Astroparticle And Plasma Physics Center (RAPP Center), Ruhr University Bochum, 44801 Bochum, Germany
\and Department of Space, Earth and Environment, Chalmers University of Technology, SE-412 96 Gothenburg, Sweden
\and Max Planck Institute for Radio Astronomy, 53121 Bonn, Germany
\and Department of Physics \& Astronomy, University of Alabama, Tuscaloosa, AL 35487, USA
\and Department of Physics, University of Wisconsin, Madison, WI 53706, USA
\and Department of Astronomy, University of Geneva, 1205 Geneva, Switzerland
\and Instituto de Estudios Astrof\'isicos, Facultad de Ingenier\'ia y Ciencias, Universidad Diego Portales, Av. Ej\'ercito Libertador 441, Santiago, Chile}

   \date{Received xxxxx}

\abstract
{}
{We present a multimessenger case study of the blazar GB6~J1542+6129, examining whether its multiwavelength behavior is consistent with neutrino production in a compact region near the central black hole, or with the parsec-scale radio jet.}
{We perform a multimessenger analysis combining $\sim$17 years of \textit{Fermi}-LAT $\gamma$-ray data, including a 5\% adaptively binned light curve and Bayesian block decomposition, with $\sim$14 years of VLBI/MOJAVE observations to derive the Doppler factor evolution of the radio core. These are compared to the temporal properties of a suspected IceCube neutrino flare with a duration of $147^{+110}_{-25}$ days, enabling a direct test of spatial and causal connections between neutrino and electromagnetic emission regions.}
{We find that the suspected neutrino flare appears to precede both a $\gamma$-ray flare and a pronounced increase in the VLBI core Doppler factor by up to $\sim1$ year. The duration of the post-flare $\gamma$-ray activity is comparable to that of the neutrino flare, which, in our framework, is consistent with both signatures originating from a single propagating disturbance whose temporal structure is preserved during the propagation, rather than being set by the light-crossing time of any single emission zone. The $\gamma$-ray spectral energy distribution remains consistent in shape across the full, flare, and post-flare intervals, indicating stable particle acceleration conditions. The temporal ordering, taken at face value, places the neutrino production site upstream of the VLBI core.}
{The observations of GB6~J1542+6129 are consistent with a disturbance-driven, multi-zone scenario in which neutrinos are produced in a compact, photon-rich inner region upstream of the parsec-scale VLBI core, plausibly at the coronal region, while the same disturbance later enhances Doppler-boosted leptonic emission once it reaches the radio core. While the data alone do not strictly establish this scenario on a population level, they show how time-domain multimessenger observations of a single AGN can localize neutrino emission relative to the parsec-scale radio jet in a particular source.}


   \maketitle

\section{Introduction}

Since the discovery of the diffuse astrophysical neutrino flux by IceCube \citep{IceCube2013}, identifying its sources has become a central challenge of multimessenger astrophysics. Active galactic nuclei (AGN), powered by accreting supermassive black holes, are among the promising candidates for the production of cosmic neutrinos \citep[e.g.][]{Begelman1984}. Their compact central regions host dense radiation fields, while relativistic jets in radio-loud AGN provide additional sites for particle acceleration. However, the dominant physical conditions and locations of neutrino production within AGN remain unresolved. Discovery-level associations, such as with the blazar TXS~0506+056 and the Seyfert galaxy NGC~1068, demonstrate that both radio-loud and radio-quiet AGN can produce detectable high-energy neutrinos \citep{ICTXS2018,ngc1068_2022}.

Effective high-energy neutrino production is expected to occur in compact regions close to the central black hole, where dense radiation fields, e.g. from the accretion disk, broad-line region, and X-ray corona, provide abundant target photons for proton--photon ($p\gamma$) interactions. Relativistic protons, accelerated to PeV--EeV energies by shocks, magnetic reconnection, or shear flows in the jet base, interact
with these ambient photons predominantly via the $\Delta^{+}$ resonance, $p + \gamma \rightarrow \Delta^{+} \rightarrow n\,\pi^{+}$ or $p\,\pi^{0}$. Charged pions decay as $\pi^{+} \rightarrow \mu^{+} + \nu_{\mu}$ followed by $\mu^{+} \rightarrow e^{+} + \nu_{e} + \bar{\nu}_{\mu}$, yielding
high-energy neutrinos that each carry $\sim$5\% of the parent proton energy, while neutral pions decay into photon pairs
($\pi^{0} \rightarrow 2\gamma$) that initiate the accompanying
electromagnetic cascade. In such environments, high-energy $\gamma$ rays are efficiently absorbed \citep{2021ApJ...911L..18K,2023A&A...679A..46K} and reprocessed to lower energies, while neutrinos escape unattenuated \citep[e.g.][]{PhysRevLett.66.2697,2016PhRvD..94j3006M,2022ApJ...941L..17M}. This naturally links neutrino emission to intrinsic hard X-ray radiation tracing compact photon fields. Observational evidence for a hard X-ray--high-energy neutrino correlation was first established in three Seyfert galaxies (NGC~1068, NGC~3079, NGC~4151) by \citet{Neronov2024}, and later extended to a broader population including Seyferts (e.g.\ CGCG 420-015) and blazars such as TXS~0506+056 and GB6~J1542+6129 \citep{Kun2024}.

In this work, we present a multimessenger analysis of GB6~J1542+6129, combining \textit{Fermi}-LAT $\gamma$-ray variability, VLBI-based Doppler factor evolution, and IceCube constraints. The blazar GB6~J1542+6129 (z=0.507) is a candidate neutrino emitter. Time-integrated searches in 10 years of IceCube data report
significances for GB6 J1542+6129 of $p \approx 0.0018$
(pre-trial) when using the 2008--2018 data set, which includes
partial-detector data from 2008--2011 \citep{IC2020tenyears}, and
$p \approx 0.0126$ when restricted to the full-detector
data collected between 2011 and 2020 \citep{ngc1068_2022}. The
IceCube multiflare analysis identifies a flare peaking on
2016-12-18 with duration $147^{+110}_{-25}$ days at a pre-trial
significance of $p \approx 0.0021$
\citep{icecubemultiflare}. Across these searches, GB6 J1542+6129
ranks as the fourth strongest neutrino association in the IceCube
source list. Together with its position on the proposed hard X-ray--neutrino luminosity relation \citep{Kun2024}, this makes GB6 J1542+6129 a prime target for studying whether the multimessenger behavior of a candidate neutrino blazar is consistent with neutrino production in a compact, photon-rich inner region.

\section{Fermi-LAT gamma-ray light curve}

We analyzed $\sim$17\,yr of \textit{Fermi}-LAT data of GB6~J1542+6129, covering the time interval from 2008 August to 2025 June \citep{Fermi4fgldr12020}. Events were selected in the energy range 100\,MeV--800\,GeV within a $20^\circ$ region of interest centered on the source position. Standard quality cuts were applied, including \texttt{evclass=128} (P8R3 SOURCE class), \texttt{evtype=3} (FRONT+BACK events), and a zenith angle cut at $90^\circ$ to suppress contamination from the Earth limb. The analysis was performed using a binned likelihood approach within the \texttt{fermiPy} framework, incorporating the standard Galactic and isotropic diffuse background models as well as cataloged point sources in the region. The minimum detection threshold to include new point sources in the model of the region of interest was TS=16. Initial light curves were produced using fixed time binning (28, 56 and 84 days), revealing significant variability but also a fraction of low-significance ($\mathrm{TS} < 16$) data points.

\begin{figure*}
\centering
\includegraphics[width=0.8\textwidth]{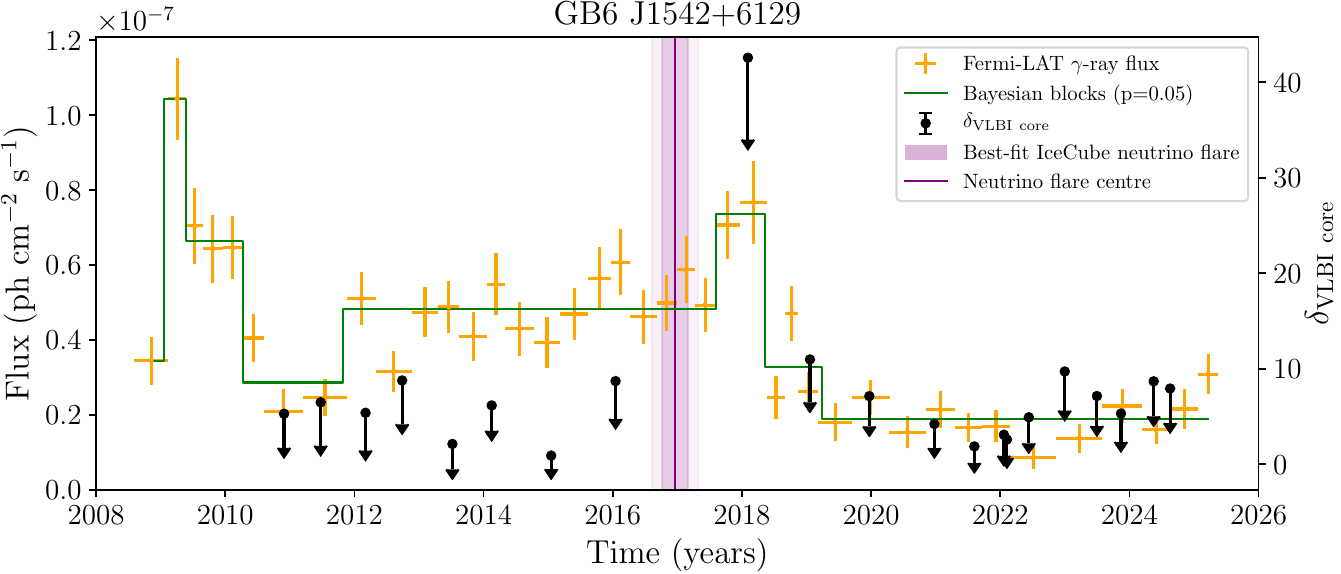}
\caption{
Multimessenger temporal evolution of the blazar GB6~J1542+6129. The orange points (left axis) show the adaptively binned 
\textit{Fermi}-LAT $\gamma$-ray flux (100\,MeV--800\,GeV), constructed 
with a constant 5\% relative uncertainty criterion. The green lines 
show the Bayesian blocks with $p=0.05$. The black points (right axis) show the Doppler factor of the VLBI core
derived from 15\,GHz MOJAVE observations. Under the equipartition
assumption (Section \ref{section:vlbi}), these are upper limits, indicated by downward
arrows, and should not be read as direct measurements; the data are
formally consistent with a constant Doppler factor. Note also the absence
of VLBI epochs between early 2016 and 2018, which prevents a precise
localization of the epoch of peak core brightness. The vertical purple line marks the central epoch 
of the suspected IceCube neutrino flare \citep{icecubemultiflare}, and the shaded regions indicate 
its central duration of $147$\,d (darker band) and its $1\sigma$ upper 
extent of $257$\,d (lighter band). The observed temporal offset 
indicates that the neutrino emission occurs earlier than the 
subsequent electromagnetic activity.}
\label{fig:gb6_multimessenger}
\end{figure*}

To obtain a statistically robust representation of the source variability, we constructed an adaptively binned light curve following a constant relative flux uncertainty criterion of 5\%. In this method, the widths of the time bins are adjusted so that each bin satisfies $\Delta F / F \approx 0.05$ in an initial, approximate likelihood fit, ensuring approximately uniform statistical significance across the light curve. This approach maximizes temporal resolution during high-flux states while preserving sensitivity during low-flux periods. Uncertainties in the flux measurements are obtained directly from the likelihood profile and correspond to statistical (1$\sigma$) errors. Fluxes in each bin were derived from likelihood fits assuming a power-law spectral model, and only bins with reliable convergence were retained. We show the resulting \textit{Fermi}-LAT $\gamma$-ray light curve of GB6~J1542+6129 in Fig. \ref{fig:gb6_multimessenger}. The resulting adaptively binned light curve reveals pronounced gamma-ray flaring activity, with enhanced emission episodes following the neutrino flare epoch.

\begin{figure*}
\centering
\includegraphics[width=0.77\textwidth]{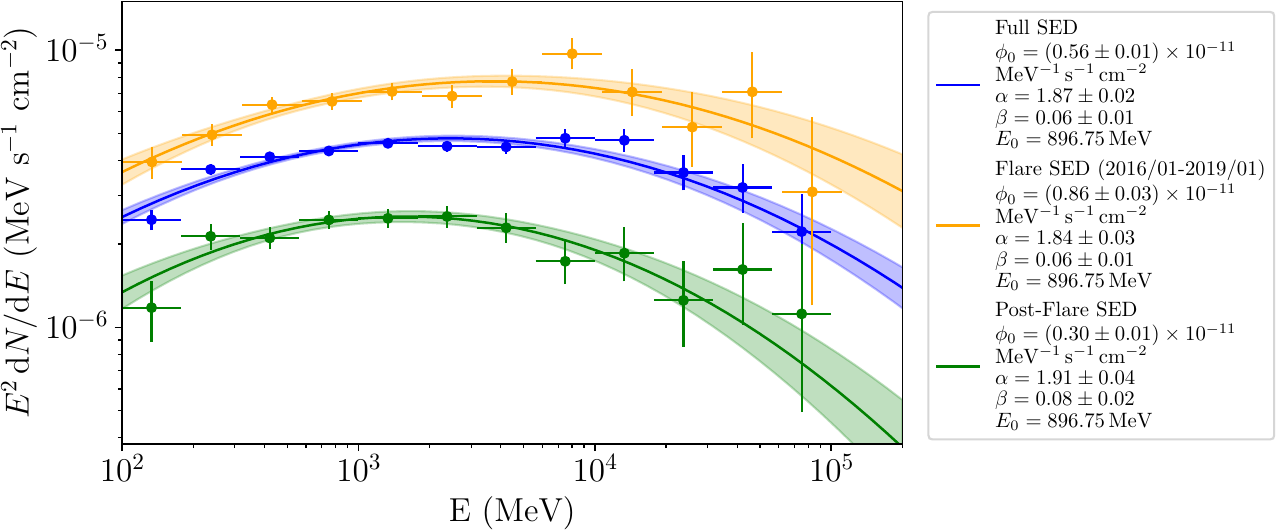} \caption{Broadband spectral energy distributions (SEDs) of GB6 J1542+6129 during the full observational period (blue), during the neutrino-associated flare interval (orange), and post-flare (green). The data points represent the \textit{Fermi}-LAT measurements in logarithmic energy bins, while the solid curves show the best-fit log-parabola models.}
\label{fig:gb6_gsed}
\end{figure*}

\section{Analysis of the archival VLBI data}
\label{section:vlbi}

We analyzed archival 15.368\,GHz Very Long Baseline Array (VLBA) data from the MOJAVE survey \citep{2018ApJS..234...12L}, covering $\sim$14\,yr of observations of GB6~J1542+6129. The calibrated visibility data were self-calibrated and imaged with \texttt{DIFMAP} \citep[][]{1997ASPC..125...77S}, and the brightness distribution was modeled with circular Gaussian components fitted to the visibilities. The innermost component was identified as the VLBI core.

For each epoch, we derived the flux density $S_\nu$, sky position and angular size $d$ of the jet component. The Doppler factor of the VLBI core $\delta$ was estimated from its brightness temperature,
\begin{equation}
T_{\mathrm{b,VLBI}} = 1.22 \times 10^{12} (1+z)\,\frac{S_\nu}{d^2 \nu^2} \ \mathrm{K}.
\end{equation}
Following \citet{1994ApJ...426...51R}, we adopt the equipartition value $T_{\mathrm{eq}} = 5\times10^{10}\,\mathrm{K}$, which corresponds to the
state in which the energy densities of the radiating relativistic electrons and the magnetic field are comparable. This choice is motivated by both theoretical and observational arguments: synchrotron sources tend to evolve toward equipartition because departures from it are energetically unfavorable and short-lived \citep{1994ApJ...426...51R}, and statistical studies of large VLBI samples find intrinsic brightness temperatures clustered close to this
value once Doppler boosting is accounted for
\citep[e.g.][]{2006ApJ...642L.115H,2018ApJ...866..137L}. We caution, however, that this assumption is best motivated for the typical (long-term, quiescent) brightness rather than the flaring state. During active phases, blazar cores can depart from equipartition toward higher intrinsic brightness temperatures, up to the inverse-Compton limit $T_{\rm IC} \sim 5\times 10^{11}$~K \citep{1969ApJ...155L..71K}.

We show the evolution of the Doppler factor of the VLBI core of GB6~J1542+6129 in Fig.\,\ref{fig:gb6_multimessenger}. The evolution of the Doppler factor is characterized by moderate values ($\delta \sim 1$--$10$) over most epochs, but exhibits a pronounced peak at epoch 2018.092, reaching $\delta_{\mathrm{core}} = 42.6 \pm 8.6$ under the equipartition assumption, with core size above the minimum resolvable size. As explained above, the peak value should therefore be regarded as an upper limit on the Doppler factor. Uncertainties on $\delta$ were derived via error propagation, accounting for measurement errors on $S_\nu$ and $d$, as well as the uncertainty introduced by the resolution limit. We note that only a single MOJAVE epoch falls within the flaring interval, which limits the temporal resolution at which the Doppler-factor evolution can be reconstructed across the flare.

\section{Discussion}

We combined \textit{Fermi}-LAT $\gamma$-ray observations with VLBI measurements of the jet structure to investigate the origin of high-energy emission in GB6~J1542+6129. The 5\% adaptively binned $\gamma$-ray light curve and the time evolution of the VLBI core Doppler factor are shown in Fig.~\ref{fig:gb6_multimessenger}. The VLBI core marks the surface where the jet becomes optically thin at the observing frequency and is typically treated as stationary.

Changes in the brightness temperature could in principle arise from variations in jet orientation, but the VLBI data disfavor this: the jet direction remains stable, with no clear outward motion or structural evolution, and the core flux density rises concurrently with the apparent Doppler factor. Since the total radio emission is strongly core-dominated and the jet ridge line is unchanged, we interpret the enhancement as a transient brightening of the VLBI core associated with the passage of a disturbance, reflecting some combination of enhanced Doppler boosting, increased emissivity, and a possible intrinsic departure from equipartition during the active state, rather than a geometric viewing-angle swing.

The multimessenger data show a temporal ordering: the suspected neutrino flare appears to precede both the $\gamma$-ray enhancement and the VLBI Doppler-factor peak by up to $\sim$1 year. We emphasize that the available data, in particular the limited cadence of the MOJAVE observations across the flare, do not allow us to measure the $\gamma$-ray--radio delay (or the neutrino--radio delay) with precision; our argument rests on the ordering of the neutrino flare center relative to the peak of the EM activity, not on a precise delay measurement. In particular, no VLBI epochs are available between early 2016
and 2018 (Fig.~\ref{fig:gb6_multimessenger}), so the epoch of peak core
brightness is only loosely localized: the implied neutrino-to-radio
ordering could range from the $\sim$1\,yr quoted here down to zero, or
even to slightly negative values of order half a year. We therefore
treat the $\sim$1\,yr figure as an indicative upper estimate of the
ordering rather than a measured delay. This sampling limitation applies specifically to the neutrino-to-radio
ordering. The $\gamma$-ray flare itself is well localized to
$\sim$2018 by the \textit{Fermi}-LAT light curve; the radio peak, by contrast,
is poorly localized because of the VLBI gap and could in principle
precede the $\gamma$-ray flare. We do not, therefore, claim an observed
radio--$\gamma$-ray simultaneity. Rather, in the synchrotron
self-Compton picture developed below the radio and $\gamma$-ray
enhancements are co-spatial and hence expected to be near-simultaneous;
this is a prediction of the scenario, which the sparse VLBI sampling can
neither confirm nor refute.

With that caveat, the $\sim$1\,year ordering is naturally interpreted as the propagation time of a disturbance, e.g. a traveling shock, magnetic reconnection event, or plasmoid, moving outward along the jet. By "disturbance" we mean a localized, transient enhancement in
the density and/or energy of the relativistic particle population,
physically realized as a travelling shock, magnetic reconnection event,
or plasmoid. At the neutrino production site in the compact inner region,
the disturbance need not yet be ultra-relativistic: it accelerates
protons close to the black hole, where the dense radiation fields make
$p\gamma$ interactions efficient, and only subsequently becomes entrained
in the fast bulk-jet flow as it propagates to parsec scales. The mildly
relativistic effective velocities inferred above are consistent with
this picture. Close to the black hole, the dense radiation fields of the accretion disk, broad-line region, and corona provide abundant target photons for $p\gamma$ interactions, so a disturbance that accelerates protons there produces neutrinos immediately, while co-spatial GeV $\gamma$-rays are absorbed via
$\gamma\gamma \rightarrow e^{+}e^{-}$ pair production on the same photon fields. As the disturbance propagates outward and the emission region becomes transparent to GeV photons, the $\gamma$-ray flare emerges with a delay set by the light-travel time between the opaque inner zone and the transparent outer jet.

In this picture, the $\gamma$-ray emission seen during the flare is the
disturbance-enhanced component, produced where the propagating
disturbance reaches the parsec-scale radio core
\citep[e.g.][]{Marscher2008,2011ApJ...726L..13A}, rather than the
steady-state inner-jet emission expected at hundreds of Schwarzschild
radii in canonical high-power blazar models
\citep[e.g.][]{2009MNRAS.397..985G}.
The radio core marks the frequency-dependent $\tau\approx1$
surface at which the jet becomes optically thin to synchrotron
self-absorption at 15\,GHz \citep{Blandford1979,Lobanov1998}. At this
location, the transiently enhanced population of relativistic electrons
produces the radio flare, and part of the same synchrotron photon field
is up-scattered to GeV $\gamma$-ray energies via the synchrotron
self-Compton process. The radio and $\gamma$-ray enhancements are
therefore expected to be essentially co-spatial, both originating at
the radio core: once the region is optically thin to the 15\,GHz radio
photons, it is also transparent to the resulting GeV $\gamma$-rays. The
spatial separation in our scenario is thus between the upstream neutrino
production site (the compact, $\gamma$-obscured inner region) and the
downstream radio/$\gamma$-ray zone at the parsec-scale core, rather than
between the radio and $\gamma$-ray emission themselves. Internal shocks
\citep{1994ApJ...421..153S,2001MNRAS.325.1559S} and magnetic
reconnection in magnetically dominated jets
\citep[e.g.][]{2016MNRAS.462...48S,2019ApJ...880...37P} are both
established mechanisms for producing such disturbances.

The disturbance path $\Delta r$, the observed delay
$\Delta t_{\rm obs}$, and the disturbance velocity $\beta_{\rm shock}$
along the jet are related by
\begin{equation}
\Delta t_{\rm obs} = (1+z)\,\frac{\Delta r}{c}\,
\left(\frac{1}{\beta_{\rm shock}} - \cos\theta\right).
\end{equation}
The VLBI core at cm wavelengths typically sits at
$\sim 10^{3}$--$10^{6}$ gravitational radii from the central black
hole \citep{Marscher2008,Hada2011}, i.e.\ $\Delta r \sim 0.025$--$25$\,pc
for an FSRQ-class blazar with $M_{\rm BH} \sim 5 \times 10^{8}\,M_\odot$ \citep[e.g.][]{2010MNRAS.402..497G,2024MNRAS.529.3699Z}.
Combined with $\Delta t_{\rm obs} \sim 1$\,yr and a small viewing
angle $\theta \sim 1/\Gamma_{\rm jet}$, this implies an effective disturbance velocity $\beta_{\rm shock}
\sim 0.1$--$1$ across the geometric range. We stress that this
range is essentially unconstrained by the present data: it excludes
only a static or strongly sub-relativistic disturbance, and is
consistent with any mildly-to-highly relativistic propagation speed.
A meaningful velocity constraint would require denser VLBI sampling
across the flare. The mere existence of an observed delay is nonetheless informative: a strictly luminal disturbance viewed head-on would produce no time difference between the upstream neutrino emission and the downstream $\gamma$-ray flare, since both would propagate at the speed of light from the same starting point.

The broadband SEDs during the full observational period, the
neutrino-to-jet active interval (2016 January 16--2019 January 19), and the post-flare interval (2019--2025) are shown in
Fig.~\ref{fig:gb6_gsed}. All three share a common log-parabola
shape within the uncertainties on $\alpha$ and $\beta$, with only
the normalization rising during the flare and dropping afterwards, consistent with a stable emission region whose brightening reflects an increase in particle or energy density rather than spectral evolution. The comparable neutrino and $\gamma$-ray durations further suggest that the observed timescale traces the lifetime of the disturbance, not the light-crossing time of a stationary zone. After the flare, the VLBI core Doppler factor returns to its pre-flare baseline while the $\gamma$-ray flux settles at a systematically lower level, indicating that the long-term emission
is governed by plasma conditions rather than boosting, consistent
with the shock partially depleting the downstream radiating
population. The disturbance may also propagate through a
pre-conditioned jet shaped by prior activity, producing
intermittent, highly non-uniform acceleration reminiscent of
lightning-like processes in turbulent plasmas
\citep{2024APh...16102976A}.

\section{Conclusion}

We presented a multimessenger analysis of the high-energy neutrino source candidate blazar GB6~J1542+6129, combining \textit{Fermi}-LAT $\gamma$-ray observations, VLBI-based Doppler factor measurements, and IceCube neutrino data (see Fig.~\ref{fig:gb6_multimessenger}). Taken at face value, the data show a temporal ordering in which the suspected neutrino flare precedes the subsequent electromagnetic activity by $\sim$1 year. This delay, together with the comparable durations of the neutrino and $\gamma$-ray flares, is consistent with a single propagating disturbance whose temporal structure is preserved during transit from a compact inner emission region to the parsec-scale jet. In this picture, the delayed $\gamma$-ray emission, closely tracked by the Doppler-factor variability, is most naturally explained as Doppler-boosted leptonic radiation enhanced by the passage of the disturbance, while the neutrino emission traces hadronic processes in a compact, $\gamma$-obscured inner region.

The observations are therefore consistent with a two-zone scenario in which a propagating disturbance produces neutrinos upstream and subsequently generates Doppler-boosted leptonic emission (plausibly including SSC and external Compton) as it propagates downstream. This framework naturally accommodates the observed time ordering, the correlation between $\gamma$-ray flux and Doppler factor (Fig.~\ref{fig:gb6_multimessenger}), and the SED stability across activity states (Fig.~\ref{fig:gb6_gsed}). Combined with the geometric corona-to-VLBI-core scale ($\sim 0.025$--$25$\,pc for $M_{\rm BH} \sim 5 \times 10^8\,M_\odot$), the observed delay is consistent with a (sub)relativistic propagating disturbance, consistent with neutrino production in the coronal region, with the disturbance subsequently propagating outward through the inner jet to the parsec-scale VLBI core. The hard X-ray--neutrino luminosity relation of \citet{Kun2024}, established on a broader AGN sample, provides empirical support for an origin in inner, radiatively efficient regions. We emphasize that, given the modest individual significance of the IceCube excess on this source and the limited cadence of the VLBI data during the flare, our results constrain a scenario consistent with inner-zone neutrino production and disturbance-driven jet variability, rather than establishing it. GB6~J1542+6129 nevertheless illustrates how time-domain multimessenger observations can be used to localize the neutrino production site within an AGN.

\begin{acknowledgements}
      We thank the anonymous referee for constructive feedback that improved this work. E.K. thanks funding from the NKFIH excellence grant TKP2021-NKTA-64. A. G. thanks support from the Pannónia Scholarship Programme. E.K., J.T. and A.F. acknowledge support from the German Science Foundation DFG, via the Collaborative Research Center \textit{SFB1491: Cosmic Interacting Matters -- from Source to Signal}. CR acknowledges support from SNSF Consolidator grant F01$-$13252 and ANID BASAL project FB210003.  This paper makes use of publicly available Fermi-LAT data provided online by the Fermi Science Support Center. We acknowledge the use of the HUN-REN Cloud. This research has made use of data from the MOJAVE database that is maintained by the MOJAVE team \citep{2018ApJS..234...12L}. The National Radio Astronomy Observatory is a facility of the National Science Foundation operated under cooperative agreement by Associated Universities, Inc. 
\end{acknowledgements}

\end{document}